\documentclass[english,superscriptaddress,aps,preprint,nofootinbib]{revtex4-2}
\usepackage[T1]{fontenc}
\usepackage{color}
\usepackage{amsmath}
\usepackage{amssymb}
\usepackage{graphicx}
\usepackage{esint}

\makeatletter
\usepackage{amsfonts}

\usepackage{babel}

\makeatother

\usepackage{babel}

\begin{document}

\title{Strong Coupling Expansion of the Entanglement Entropy of Yang-Mills
Gauge Theories }

\author{Jiunn-Wei Chen}
\email{jwc@phys.ntu.edu.tw}
\affiliation{Department of Physics and Center for Theoretical Sciences, National
Taiwan University, Taipei 10617, Taiwan}
\affiliation{Leung Center for Cosmology and Particle Astrophysics, National Taiwan
University, Taipei 10617, Taiwan}
\affiliation{Center for Theoretical Physics, Massachusetts Institute of Technology,
Cambridge, MA 02139, USA}

\author{Shou-Huang Dai}
\email{shdai@stust.edu.tw}
\affiliation{Center for General Education, Southern Taiwan University of Science
and Technology, Tainan 71005, Taiwan}

\author{Jin-Yi Pang}
\email{axial.pang@gmail.com}
\affiliation{University of Shanghai for Science and Technology, Shanghai 200093,
PR CHINA}
\affiliation{Department of Physics and Center for Theoretical Sciences, National
Taiwan University, Taipei 10617, Taiwan}

\begin{abstract}
We propose a novel prescription for calculating the entanglement entropy
of the $SU(N)$ Yang-Mills gauge theories on the lattice under the
strong coupling expansion in powers of $\beta=2N/g^{2}$, where $g$
is the coupling constant. Using the replica method, our Lagrangian
formalism maintains gauge invariance on the lattice. At $O(\beta^{2})$
and $O(\beta^{3})$, the entanglement entropy is solely contributed
by the central plaquettes enclosing the conical singularity of the
$n$-sheeted Riemann surface. The area law emerges naturally to the
highest order $O(\beta^{3})$ of our calculation. The leading $O(\beta)$
term is negative, which could in principle be canceled by taking into
account the ``cosmological constant'' living in interface of the
two entangled subregions. This unknown cosmological constant resembles
the ambiguity of edge modes in the Hamiltonian formalism. We further
speculate this unknown cosmological constant can show up in the entanglement
entropy of scalar and spinor field theories as well. Furthermore,
it could play the role of a counterterm to absorb the ultraviolet
divergence of entanglement entropy and make entanglement entropy a
finite physical quantity. 
\end{abstract}

\maketitle

\section{Introduction}

Entanglement entropy is a measure of the level of entanglement between
the degrees of freedom in two subregions of a physical system. Besides
being a fundamental and mysterious property of quantum mechanics and
quantum field theory, quantum entanglement is of practical use. In
some systems, the entanglement entropy plays the role of an order
parameter characterizing the quantum phase transitio\textcolor{black}{ns,
and de}monstrates the scaling behavior near the critical point \cite{Klebanov:2007ws,Osborne2002,PhysRevA.81.032304,Chen:2014wka,Vidal2003,Latorre2004}.
In field theory, a widely used method for calculating the entanglement
entropy is the replica trick \cite{Callan:1994py}. This method calculates
the trace of the reduced density matrix to the $n$-th power in the
path integral formalism, which amounts to computing the free energy
of the system on a $n$-sheeted Riemann surface, or equivalently on
a ``cone'' with a conical angle of $2n\pi$. The entanglement entropy
is then obtained as a response of the free energy to the change of
the conical angle at $n=1$. This is similar to calculating the black
hole entropy by evaluating the response of the free energy of fields
with respect to the deficit angle in the Euclidean spacetime \cite{Kabat:1995eq}.

In previous studies, while the computation of entanglement entropy
for the scalar and the spinor fields are considered straightforward,
it becomes more subtle for the gauge fields. It was found by Kabat
\cite{Kabat:1995eq} that the gauge fields in the black hole entropy
method yields an extra negative contact term compared to their entanglement
entropy. This term is due to the interaction of the particles with
the horizon, and is believed to be related to the choice of the boundary
condition while removing the tip of the cone due to the black hole
in the Euclidean space. It was later interpreted by \cite{Donnelly:2012st}
as arising from incorrect treatment of the zero modes.

The ambiguity in calculating the entanglement entropy of the gauge
fields is caused by that the global Hilbert space of gauge invariant
physical state does not admit a direct product decomposition between
the regions on the two sides of the entangling surface. To deal with
this issue, \cite{Aoki:2015bsa,Ghosh:2015iwa,Soni:2016ogt} propose
to embed the original Hilbert space in an extended Hilber space allowing
a tensor product factorization.

On the other hand, the entanglement entropy of the gauge theories
had been studied with the Hamiltonian \cite{Donnelly:2014fua,Donnelly:2011hn,Casini:2013rba,Buividovich:2008gq,Ghosh:2015iwa,Buividovich:2008yv,Radicevic:2015sza}
and the Lagrangian \cite{Velytsky:2008rs,Buividovich:2008kq,Gromov:2014kia,Agarwal:2016cir}
approaches. In the Hamiltonian approach for the case of the gauge
fields on the lattice \cite{Donnelly:2014fua,Donnelly:2011hn,Casini:2013rba,Buividovich:2008gq},
one needs to impose the Gauss law or the gauge fixing constraints
in order to get rid of the unphysical degrees of freedom\textcolor{black}{.}
This leads to the difficulty in decomposing the global gauge invariant
states into the direct products of those living in each subregion.
It was proposed \cite{Donnelly:2014fua} that these ambiguities might
be compensated by edge modes living in the interface of the subregions
and determined by the transverse electric fields. This echoes the
concept that the contact term in \cite{Kabat:1995eq} arises from
the sources on the horizon. In \cite{Donnelly:2014fua} this negative
contact term was found to arise from the entanglement of the edge
modes. Moreover, in order to clarify the issue of the Hilbert space
decomposition, \cite{Casini:2013rba} employs various choices of the
electric, the magnetic, and the trivial centers in the operator algebra
in the Hamiltonian formalism. The electric and magnetic choices also
result in different tripartite topological entanglement entropy. By
using Hamiltonian formalism dominated by the electric term, \cite{Radicevic:2015sza}
shows that for the $SU(N)$ lattice gauge theory in the strongly coupled
regime, the entanglement entropy behaves as $|\partial V_{\perp}|\frac{\text{log }g^{2}N}{g^{8}N^{2}}$.

In view of these ambiguities, we use a complementary approach --
the Lagrangian formalism -- to shed light on the entanglement entropy
of the Yang-Mills theory on the lattice from a different perspective.
We apply the replica method and take the derivative of the trace of
$n$ copies of the reduce density matrix by $n$. Since the density
matrix \footnote{We remind the readers that, here and in the following, by ``the density
matrix'' we mean ``the ground state density matrix at strong coupling''.} can always be expressed in the path integral formalism, the replica
method is valid for the gauge fields at continuum. We start from the
replica method at continuum, and then discretize the spacetime into
a hypercubic lattice. We divide an infinitely large system into two
semi-infinite subregions by a flat boundary, and decompose the spacetime
into a direct product of a 2-dimensional cone with a conical angle
of $2n\pi$ (or equivalently an $n$-sheeted Riemann surface) and
an ordinary Euclidean space transverse to the cone. (See Section \ref{sec:Caculation}
for the detail.) Then we use the Wilson gauge action on a discrete
lattice. This action sums over the Wilson loops on the plaquettes,
including those on the $n$-sheeted Riemann surface and those on the
ordinary Euclidean space. In contrast with the previous studies where
the conical singularity is placed on the lattice site, we use a different
discretization setup by locating the conical singularity in the center
of the plaquette. As a result, the branch lines on the $n$-sheeted
Riemann surface cut across the links, and there is no lattice site
on the cut. (See Fig. \ref{fig:The--sheeted-manifold}.) This setup
yields two types of plaquettes on the lattice of the cone: the central
plaquettes encircling the tip of the cone, formed by $4n$ links,
and the regular plaquettes formed by 4 links with no singularity inside.
When the conical angle, or $n$, changes, only the central plaquettes
(i.e. those plaquettes with $4n$ links) respond to this change. As
a result, entanglement entropy necessarily involves those central
plaquettes. The fact that all of the central plaquettes live across
the interface between the two subregions naturally give rise to the
area law, which states that the leading contribution to entanglement
entropy scales as the area of the interface.

The connection to the area law can be further demonstrated order by
order diagrammatically under the strong coupling expansion (see for
example \cite{Balian:1974xw}). Interestingly, we find the leading
term in the strong coupling expansion negative. However, symmetries
of the action allow a two-dimensional cosmological constant living
in the interface \cite{Cooperman:2013iqr} which could provide a positive
contribution at an even lower order. We speculate that the freedom
to tune this unknown 2-dimensional cosmological constant corresponds
to the ambiguities encountered in the Hamiltonian approach.\textbf{\ }We
further speculate that the 2-dimensional cosmological constant can
show up in the entanglement entropy of scalar and spinor field theories
as well. They can play the role of a counterterm to absorb the ultraviolet
divergence of entanglement entropy and make entanglement entropy a
finite physical quantity.

This paper is organized as follows. In Sec. II, we briefly review
the notion of entanglement entropy and the replica method. The entanglement
entropy of Yang-Mills fields on the lattice under the strong coupling
expansion is calculated in Sec. III, and the cancellation of the negative
contribution by including a 2-dimensional cosmological constant is
discussed in Sec. IV. Sec. V compares our result with the previous
ones obtained by the Hamiltonian methods, and Sec. VI summarizes our
study.

\section{Entanglement Entropy and the Replica Method}

Suppose our system occupies an infinitely large and flat $d+1=4$
dimensional spacetime. The $d=3$ dimensional space is divided into
two regions $A$ and $B$ by a time independent, infinite, and flat
2-dimensional space-like boundary. The entanglement entropy (EE) of
a quantum theory between the two subregions is defined by the von
Neumann entropy. With some simple algebra, it can be re-expressed
as: 
\begin{equation}
S_{\text{EE}}=-\text{tr}[\rho_{A}\ln\rho_{A}]=-\left.\frac{\partial}{\partial n}\right\vert _{n\rightarrow1}\ln\text{tr}[\rho_{A}^{n}]
\end{equation}
where $\rho_{A}=\text{tr}_{B}[\rho]$ is the reduced density matrix
by tracing out the degrees of freedom in region $B$. This expression
is called the replica method because it involves $n$ copies of $\rho_{A}$.

\begin{figure}
\begin{centering}
\includegraphics[scale=0.5]{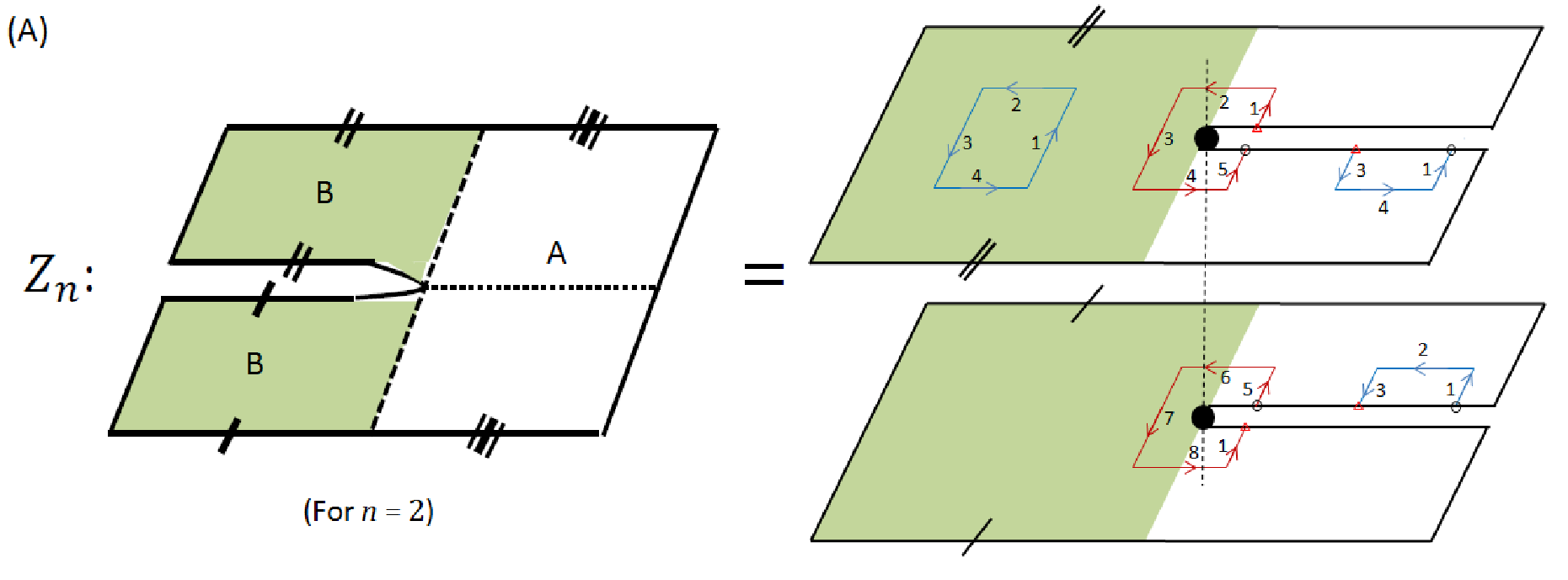} 
\par\end{centering}
\begin{centering}
\includegraphics[scale=0.5]{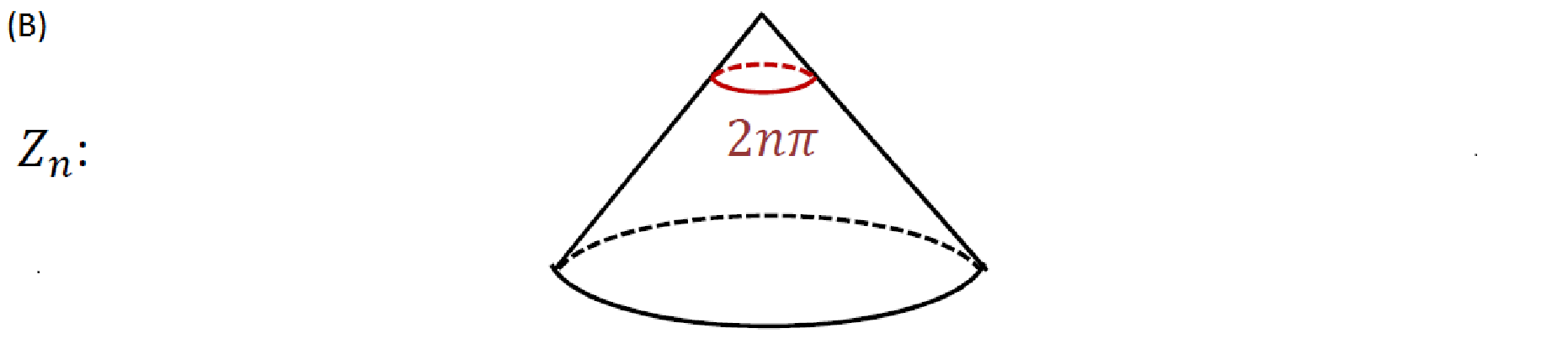} 
\par\end{centering}
\protect\caption{The $n$-sheeted Riemann surface in the replica trick. (A) illustrates
the equivalent geometric structures of the n-sheeted manifold in the
case of $n=2$ arizing from the replica trick, on which the partition
function $Z_{n}$ is computed. The unshaded and the shaded parts represent
the subregion A and B, respectively. The subregion B is traced out
in the reduced density matrix $\rho_{A}$. The 2 sheets on the right
are on top of each other at the same $x_{\perp}$ coordinates, attaching
to each other at the conical singularity across the branch cut. The
conical sigularity is represented by the black dot and the dotted
vertical line. In the right figure, it is also demonstrated explicitly
that, after discretization, the central plaquette with $4n$ edges
encircling the conical singularity, and the ordinary plaquettes with
4 links located on a sheet or across two adjacent sheets (see also
Fig. \protect\ref{fig:plaquettes}). The numbers label the order
of the link variables forming the plaquettes. As the sizes of the
space and the Euclidean time are both infinite (i.e. $T=0$), this
geometry is equivalent to a cone with $2n\pi$ conical angle depicted
in (B).}
\label{fig:The--sheeted-manifold} 
\end{figure}

An elegant path integral formulation to compute the entanglement entropy
using the replica method was first introduced in \cite{Callan:1994py}
(see also \cite{Hertzberg:2010uv}). In this set up, one recalls that
$\rho_{ij}=\left\langle i\left\vert e^{-H/T}\right\vert j\right\rangle $
and Tr$[\rho]$ is the partition function calculated in finite temperature
field theory with appropriate boundary conditions (periodic and anti-periodic
boundary conditions for bosons and fermions, respectively) imposed
for fields at Euclidean time $\tau=0$ and $1/T$, where $T$ is the
temperature. Then Tr$[\rho^{2}]$ can be computed by doubling the
period (by imposing appropriate boundary conditions at $\tau=0$ to
$2/T$). Similarly, Tr$[\rho_{A}^{2}]$ is computed by doubling the
period (from $0$ to $2/T$) for region $A$ while maintaining the
single period (from $0$ to $1/T$) for the region $B$, as shown
in the left plot of Fig. \ref{fig:The--sheeted-manifold}(A), which
is equivalent to performing the path integral on a 2-sheeted Riemann
surface in the right plot. One can generalize this set up to Tr$[\rho_{A}^{n}]$
for an arbitrary $n$. There is no restriction on the space partition
between $A$ and $B$. The sizes of $A$, $B$, and $T$ can be either
finite or infinite.

In this paper, we will just concentrate on the simplest case with
the sizes of space and (Euclidean) time to be both infinite (i.e.
$T=0$)\ and the interface between $A$ and $B$ to be a flat infinite
plane. In this limit, the $n$-sheeted Riemann surface has a conical
structure as shown in Fig. \ref{fig:The--sheeted-manifold}(B) with
the time and longitudinal spatial direction (orthogonal to the interface
between the A, B region) lying on the cone while the space on the
interface transverse to the cone.

As a result, $\text{tr}[\rho_{A}^{n}]$ becomes a partition function
$Z_{n}$ on the $n$-sheeted Riemann surface, or, in our case, a cone
with $2n\pi$ conical angle, normalized by the $n$-copies of the
partition function on the ordinary Euclidean space, $Z_{1}^{n}$:
\begin{equation}
\text{tr}[\rho_{A}^{n}]=\frac{Z_{n}}{Z_{1}^{n}},\label{eq:replica1}
\end{equation}
which ensures that as $n=1$, $\text{tr}[\rho_{A}^{n}]$=1. The entanglement
entropy is then given by 
\begin{equation}
S_{\text{EE}}=-\left.\frac{\partial}{\partial n}\left(\ln Z_{n}-n\ln Z_{1}\right)\right\vert _{n\rightarrow1}\overset{n=1+\epsilon}{=}-\frac{1}{\epsilon}\left[\ln Z_{1+\epsilon}-(1+\epsilon)\ln Z_{1}\right]_{\epsilon\rightarrow0}.
\end{equation}

Note that in the integral of $Z_{n}$, $n$ is taken as an integer.
After the analytic expression for $\text{tr}[\rho^{n}]$ is obtained,
then $n$ can be analytically extended to non-integers to carry out
the differentiation at $n=1$.

\section{Calculation on an n-sheeted Lattice Manifold\label{sec:Caculation}}

\begin{figure}
\begin{centering}
\includegraphics[scale=0.5]{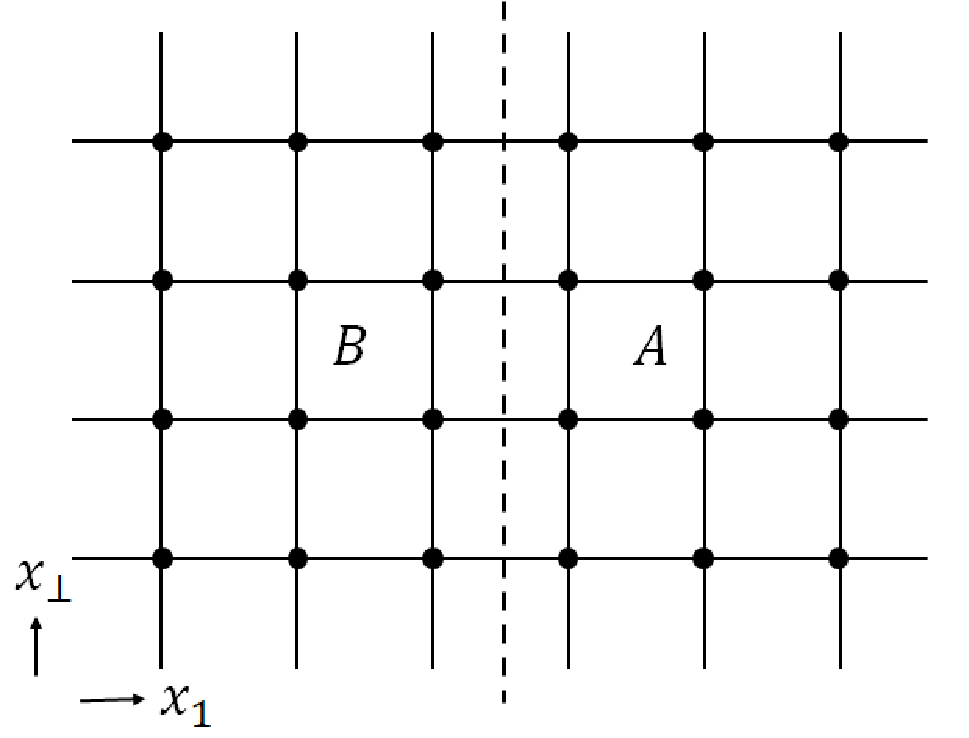}
\par\end{centering}
\protect\caption{The cartoon diagram depicting our system setup in which the boundary
between the regions A and B cuts across the links connecting the two
regions. In our coordinates, $x_{1}$ is orthogonal to the boundary,
while the boundary is along the $(x_{2},x_{3})$ dimensions which
are abbreviated as $x_{\perp}$. In this diagram, one of the $x_{\perp}$
dimensions is suppressed.}
\label{fig:setup}
\end{figure}

We now discretize the 4-dimensional spacetime by a hypercubic lattice.
The 3 spatial dimensions are labeled by the coordinates $(x_{1},x_{2},x_{3})$,
while the Euclidean time direction is labeled by $\tau$. Our system
is set up such that the flat boundary between the regions A and B
cuts across the links connecting the lattice sites in the two regions,
as shown in Fig. \ref{fig:setup}. The boundary is in the $(x_{2},x_{3})$
plane located at $x_{1}=0$.

When we calculate $Z_{n}$, the spacetime is decomposed into the direct
product of a $1+1$ dimensional $n$-sheeted Riemann lattice and an
Euclidean squared lattice spanning the other 2 dimensions. The $1+1$
dimensional $n$-sheeted Riemann surface is labeled by $(x_{1},\tau)$,
while the other two Euclidean dimensions are described by the coordinates
$(x_{2},x_{3})$. In the following sections in this paper, for our
convenience we will call $x_{1}$ by $x_{\parallel}$, and $(x_{2},x_{3})$
by $x_{\perp}$. We remind the readers that $\parallel$ and $\perp$
directions are taken with respect to the $(x_{1},\tau)$ Riemann surface,
but not to the boundary surface of the regions A, B. In the circumstances
that using the boundary as the reference plane is necessary, we will
describe explicitly in the text.

\begin{figure}
\begin{centering}
\includegraphics{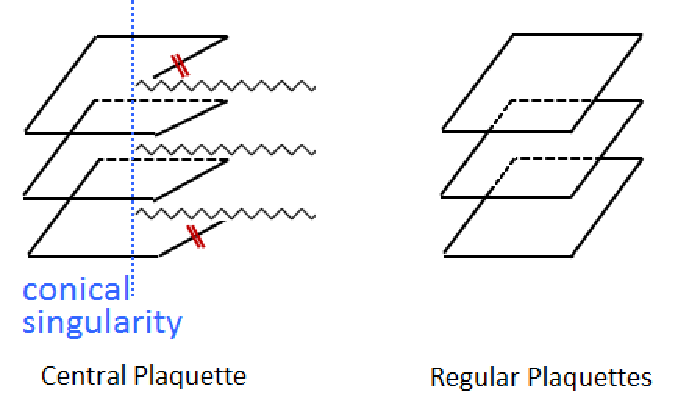} 
\par\end{centering}
\protect\caption{One central plaquette (left) enclosing the conical singularity v.s.
three regular plaquettes on three different sheets (right) in the
case of $n=3$ sheets. In the left figure, the dotted line represents
the conical singularity to which all the three sheets attach at the
same $x_{\perp}$., and the wavy lines are cuts on each sheet. The
central plaquette has 12 edges and it takes $6\pi$ to go around it,
while each regular plaquette has 4 links.}
\label{fig:plaquettes} 
\end{figure}

By means of this discritization, the end point of the branch cut on
the $n$-sheeted Riemann surface (corresponding to the conical singularity
at the tip the cone) locates inside a plaquette, rather than sitting
on a lattice site. As a result of the discretization, there are two
types of plaquettes. The \textit{ordinary plaquettes} contain no conical
singularity inside, and are composed of 4 links. They are in the transverse
$x_{\perp}$ and the parallel $(x_{\parallel},\tau)$ dimensions.
Their plaquette variables are denoted by $U_{\Box}^{(k)}=\mathcal{P}\prod_{l\in\Box}U_{l}^{(k)}$,
where $k$ indicates that they locate on the $k$-th sheet, and $\mathcal{P}$
signifies the ordered product of the link variables $U_{l}^{(k)}$
forming the plaquette. On the other hand, the \textit{central plaquette}
going around all the $n$ sheets encloses the conical singularity
on the Riemann surface in the parallel $(x_{\parallel},\tau)$ dimensions,
and are composed of $4n$ links. Their plaquette variables are expressed
by $U_{\boxdot}(x_{\perp})=\mathcal{P}\prod_{l\in\boxdot}U_{l}$,
where $x_{\perp}$ is the location of the $(x_{\parallel},\tau)$
planes in the transverse dimensions. See Figs. \ref{fig:The--sheeted-manifold}
and \ref{fig:plaquettes} for the illustration of these two types
of plaquettes. It will be clear later that our result is independent
of the location of the conical singularity as long as it is encircled
by a plaquette. For simplicity we choose that the conical singularity
sits in the center of a plaquette.

Recall that the partition function of the lattice gauge theory on
a one-sheet manifold is given by 
\begin{equation}
Z=\int\mathcal{D}U\exp\left\{ -\beta\sum_{\Box}\left[1-\frac{1}{N}\text{Re }\text{tr}U_{\Box}\right]\right\} \underset{a\rightarrow0}{\longrightarrow}\int\mathcal{D}A\exp\left\{ -\int d^{4}x\left[\frac{1}{4g^{2}}\text{tr}F^{2}\right]\right\} ,\label{eq:SGc}
\end{equation}
where $\beta=2N/g^{2}$ and $\Box$ labels the location of plaquettes.
The plaquette variable $U_{\Box}$ is the local Wilson loop composed
of the ordered product of four gauge links,\textcolor{red}{{} }$U_{\Box}=\mathcal{P}\prod_{l\in\Box}U_{l}$
where $\mathcal{P}$ indicates the ordered product and $U_{l}$ is
the link variable representing the gauge fields. The action recovers
the Yang-Mills action in the continuum limit by setting the lattice
spacing $a\rightarrow0$. We assume that the lattice has the same
spacing $a$ in all dimensions.

To construct a lattice gauge field system in the $d+1=4$ dimensions
whose $1+1$ dimensions is an $n$-sheeted manifold, we rewrite the
partition function in Eq.(\ref{eq:SGc}) in terms of the ordinary
plaquettes $U_{\Box}^{(k)}$ and the central ones $U_{\boxdot}(x_{\perp})$,
such that the partition function reads 
\begin{eqnarray}
Z_{n} & = & \int\left[\prod_{m=1}^{n}\mathcal{D}U^{(m)}\right]\exp\left\{ \frac{\beta}{N}\sum_{k=1}^{n}\sum_{\Box}\text{Re}\text{tr}[U_{\Box}^{(k)}-1]+\frac{\beta}{nN}\sum_{\mathbf{x}_{\perp}}\text{Re }\text{tr}[U_{\boxdot}(\mathbf{x}_{\perp})-1]\right\} \label{eq:partition}\\
 & \underset{a\rightarrow0}{\rightarrow} & \int\left[\prod_{m=1}^{n}\mathcal{D}A^{(m)}\right]\exp\left\{ -\frac{1}{4g^{2}}\int d^{2}x_{\perp}\Big(\sum_{k=1}^{n}\int_{\mathbb{R}^{2}-\{0\}}d^{2}x_{\parallel}\text{tr}F^{(k)2}\right.\nonumber \\
 &  & \left.\hspace{2cm}\hspace{4cm}\hspace{0.5cm}+\frac{1}{n}\sum_{k,l=1}^{n}\int_{\{0\}}d^{2}x_{\parallel}\text{tr}\left[F^{(k)}F^{(l)}\right]\Big)\right\} \nonumber \\
 & = & \int\left[\prod_{m=1}^{n}\mathcal{D}A^{(m)}\right]\exp\left\{ -\frac{1}{4g^{2}}\sum_{k=1}^{n}\int d^{2}x_{\perp}\int d^{2}x_{\parallel}\text{tr}F^{(k)2}\right\} ,\nonumber 
\end{eqnarray}
where the upper indices $(k)$ label different sheets on the Riemann
surface. We have assumed the boundary condition 
\begin{equation}
F^{(k)}(\mathbf{x}_{\perp},\mathbf{x}_{\parallel}=\mathbf{0})=F^{(l)}(\mathbf{x}_{\perp},\mathbf{x}_{\parallel}=\mathbf{0}),\;(k\neq l),
\end{equation}
i.e. the field strength on each sheet is assumed be equal. This condition
is natural if we impose $2\pi$ rotation symmetry on the $n$-sheeted
surface. We have introduced an extra $1/n$ factor to the central
plaquette terms, because the $4n$ gauge links of the central plaquettes
encircle the $F_{01}$\ flux over an area $na^{2}$, and give rise
to a factor $\propto n^{2}a^{4}F_{01}^{2}$ to the action, inconsistent
with the contribution from the transverse plaquettes scaling as $na^{4}$.
Therefore a factor of $1/n$ is introduced to compensate this effect.

Expanding Eq.(\ref{eq:partition}) to the second order of $\beta$,
we have 
\begin{eqnarray}
Z_{n} & = & e^{-\beta n\mathcal{N}-\frac{\beta}{n}\mathcal{N}_{\perp}}\int\left[\prod_{m=1}^{n}\mathcal{D}U^{(m)}\right]\Bigg\{1+\frac{\beta}{N}\sum_{k=1}^{n}\sum_{\Box}\text{Re }\text{tr}\left[U_{\Box}^{(k)}\right]+\frac{\beta}{nN}\sum_{\mathbf{x}_{\perp}}\text{Re }\text{tr}\left[U_{\boxdot}(\mathbf{x}_{\perp})\right]\nonumber \\
 &  & +\frac{\beta^{2}}{2N^{2}}\sum_{k,l=1}^{n}\sum_{\Box,\Box^{\prime}}\left[\text{Re }\text{tr}U_{\Box}^{(k)}\right]\left[\text{Re }\text{tr}U_{\Box^{\prime}}^{(l)}\right]+\frac{\beta^{2}}{2N^{2}}\frac{1}{n^{2}}\sum_{\mathbf{x}_{\perp1},\mathbf{x}_{\perp2}}\left[\text{Re }\text{tr}U_{\boxdot}(\mathbf{x}_{\perp1})\right]\left[\text{Re }\text{tr}U_{\boxdot}(\mathbf{x}_{\perp2})\right]\nonumber \\
 &  & +\frac{\beta^{N}}{N!N^{N}}\left[\sum_{k=1}^{n}\sum_{\Box}\text{Re }\text{tr}\left[U_{\Box}^{(k)}\right]+\frac{1}{n}\sum_{\mathbf{x}_{\perp}}\text{Re }\text{tr}\left[U_{\boxdot}(\mathbf{x}_{\perp})\right]\right]^{N}\nonumber \\
 &  & +\text{\ensuremath{\beta^{2}}(tr}U_{\Box})(\text{tr}U_{\boxdot})\text{ cross-terms}+\text{other higher order terms in}\beta\Bigg\}.\label{eq:zn1}
\end{eqnarray}
In this equation, $N$ is the number of colors. $\mathcal{N}$ is
the number of plaquettes not encircling the conical singularity on
a single sheet, including those parallel and transverse to the $(x_{\parallel},\tau)$
Riemann surface. $\mathcal{N}_{\perp}$ is the number of central plaquettes
encircling the conical singularity at $(x_{\parallel}=0,\tau=0)$
if we count across all $x_{\perp}$ coordinates, i.e. $\mathcal{N}_{\perp}=A_{\perp}/a^{2}$
where $A_{\perp}$ is the area of the boundary between the regions
A, B and $a$ is the lattice size. We isolate the $O(\beta^{N})$
terms in the third line of (\ref{eq:zn1}) explicitly from ``$\text{other higher order terms in}\beta$'',
because they will contribute to $O(\beta^{2})$ and $O(\beta^{3})$
in the $N=2,3$ cases respectively. It will be understood more clearly
in the following.

To calculate the result, we apply $\text{Re}\ \text{tr}U=\frac{1}{2}(\text{tr}U+\text{tr}U^{\dagger})$
to (\ref{eq:zn1}), and use the Harr measure

\begin{eqnarray}
\int dUU_{ij} & = & 0\;=\;\int dUU_{ij}^{\dagger},\label{eq:me1}\\
\int dUU_{ij}U_{kl}^{\dagger} & = & \frac{1}{N}\delta_{il}\delta_{jk},\label{eq:me2}
\end{eqnarray}
such that 
\begin{eqnarray}
\int\mathcal{D}U\text{tr}U_{\Box}^{(k)}\text{tr}U_{\Box}^{(l)\dagger} & = & \delta_{kl},\label{a}\\
\int\mathcal{D}U\text{ tr}U_{\boxdot}(\mathbf{x}_{\perp1})\text{ tr}U_{\boxdot}^{\dagger}(\mathbf{x}_{\perp2}) & = & \delta_{x_{\perp1}x_{\perp2}},\label{b}
\end{eqnarray}
to compute the contribution of the Wilson loops from $\int\mathcal{D}U\text{ tr}U\text{tr}U^{\dagger}$.
Note that (\ref{a}) and (\ref{b}) are independent of $n$, and $\mathcal{D}U$
in (\ref{a}) and (\ref{b}) are for integration of all link variables.
It can be derived that the $\text{(tr}U_{\Box})(\text{tr}U_{\boxdot})\text{ cross-terms}$
vanishes in (\ref{eq:zn1}). We then obtain
\begin{eqnarray}
Z_{n} & = & e^{-\beta n\mathcal{N}-\frac{\beta}{n}\mathcal{N}_{\perp}}\left\{ 1+\frac{1}{2}\frac{\beta^{2}}{2N^{2}}\sum_{k,l=1}^{n}\sum_{\Box,\Box^{\prime}}\int\left[\prod_{m=1}^{n}\mathcal{D}U^{(m)}\right]\left[\text{tr}U_{\Box}^{(k)}\right]\left[\text{tr}U_{\Box^{\prime}}^{(l)\dagger}\right]\right.\nonumber \\
 &  & \qquad+\frac{1}{2}\frac{\beta^{2}}{2N^{2}}\frac{1}{n^{2}}\sum_{\mathbf{x}_{\perp1},\mathbf{x}_{\perp2}}\int\left[\prod_{k=1}^{n}\mathcal{D}U^{(k)}\right]\left[\text{tr}U_{\boxdot}(\mathbf{x}_{\perp1})\right]\left[\text{tr}U_{\boxdot}^{\dagger}(\mathbf{x}_{\perp2})\right]\nonumber \\
 &  & \qquad+\frac{\beta^{N}}{N!N^{N}}\int\left[\prod_{m=1}^{n}\mathcal{D}U^{(m)}\right]\left[\left(\sum_{k=1}^{n}\sum_{\Box}\text{Re }\text{tr}\left[U_{\Box}^{(k)}\right]\right)^{N}+\left(\frac{1}{n}\sum_{\mathbf{x}_{\perp}}\text{Re }\text{tr}\left[U_{\boxdot}(\mathbf{x}_{\perp})\right]\right)^{N}\right]\nonumber \\
 &  & \left.\qquad+O(\beta^{4})\right\} \nonumber \\
 & = & e^{-\beta n\mathcal{N}-\frac{\beta}{n}\mathcal{N}_{\perp}}\left\{ 1+\frac{\beta^{2}}{4N^{2}}n\mathcal{N}+\frac{\beta^{2}}{4N^{2}}\frac{1}{n^{2}}\mathcal{N}_{\perp}+\frac{2}{N!}\Big(\frac{\beta}{2N}\Big)^{N}\left[n\mathcal{N}+\frac{\mathcal{N}_{\perp}}{n^{N}}\right]+O(\beta^{4})\right\} .\label{eq:zn2}
\end{eqnarray}
Because of the Harr measure (\ref{a}) and (\ref{b}), the $O(\beta^{3})$
terms in the ``other higher order terms in $\beta$'' of (\ref{eq:zn1})
vanish. They instead start from $O(\beta^{4})$ order, corresponding
to the $\text{tr}U_{\Box}\text{tr}U_{\Box}^{\dagger}\text{tr}U_{\Box}\text{tr}U_{\Box}^{\dagger}$
and $\text{tr}U_{\boxdot}\text{tr}U_{\boxdot}^{\dagger}\text{tr}U_{\boxdot}\text{tr}U_{\boxdot}^{\dagger}$
terms.

The $N$ Wilson loops interactions the third line of (\ref{eq:zn2})
is derived from the third line of (\ref{eq:zn1}), and give rise to
the $\beta^{N}$ terms in the result of (\ref{eq:zn2}). We explain
how to calculate these terms as follows. By making use of the identities
\begin{eqnarray}
\int dUU_{i_{1}j_{1}}^{(p)}U_{i_{2}j_{2}}^{(q)}...U_{i_{N}j_{N}}^{(s)} & = & \frac{1}{N!}\epsilon_{i_{1}i_{2}...i_{N}}\epsilon_{j_{1}j_{2}...j_{N}}\delta^{pq\cdots s},\\
\int dUU_{i_{1}j_{1}}^{(p)\dagger}U_{i_{2}j_{2}}^{(q)\dagger}...U_{i_{N}j_{N}}^{(s)\dagger} & = & \frac{1}{N!}\epsilon_{i_{1}i_{2}...i_{N}}\epsilon_{j_{1}j_{2}...j_{N}}\delta^{pq\cdots s},
\end{eqnarray}
the contribution from the $N$ Wilson loops interactions $\int\mathcal{D}U\left(\text{tr}U\right)^{N}$
and $\int\mathcal{D}U\left(\text{tr}U^{\dagger}\right)^{N}$of $O(\beta^{N})$
are

\begin{eqnarray}
\int\mathcal{D}U\left(\text{tr}U_{\Box}\right)^{N}=\int\mathcal{D}U\left(\text{tr}U_{\Box}^{\dagger}\right)^{N} & = & 1,\\
\int\mathcal{D}U\left[\text{tr}U_{\boxdot}(\mathbf{x}_{\perp})\right]^{N}=\int\mathcal{D}U\left[\text{tr}U_{\boxdot}^{\dagger}(\mathbf{x}_{\perp})\right]^{N} & = & 1.
\end{eqnarray}
Therefore, the $O(\beta^{N})$ contribution from these terms can be
computed,

\begin{eqnarray}
Z_{n}(\beta^{N}) & = & \int\left[\prod_{m=1}^{n}\mathcal{D}U^{(m)}\right]\frac{\beta^{N}}{N!N^{N}}\left\{ \sum_{k=1}^{n}\sum_{\Box}\text{Re }\text{tr}\left[U_{\Box}^{(k)}\right]+\frac{1}{n}\sum_{\mathbf{x}_{\perp}}\text{Re }\text{tr}\left[U_{\boxdot}(\mathbf{x}_{\perp})\right]\right\} ^{N}\nonumber \\
 & = & \int\left[\prod_{m=1}^{n}\mathcal{D}U^{(m)}\right]\frac{\beta^{N}}{N!N^{N}}\left\{ \left(\sum_{k=1}^{n}\sum_{\Box}\text{Re }\text{tr}\left[U_{\Box}^{(k)}\right]\right)^{N}+\left(\frac{1}{n}\sum_{\mathbf{x}_{\perp}}\text{Re }\text{tr}\left[U_{\boxdot}(\mathbf{x}_{\perp})\right]\right)^{N}\right\} \nonumber \\
 & = & \int\left[\prod_{m=1}^{n}\mathcal{D}U^{(m)}\right]\frac{\beta^{N}}{N!(2N)^{N}}\left\{ \left(\sum_{k=1}^{n}\sum_{\Box}\text{tr}\left[U_{\Box}^{(k)}\right]\right)^{N}+\left(\sum_{k=1}^{n}\sum_{\Box}\text{tr}\left[U_{\Box}^{(k)\dagger}\right]\right)^{N}\right.\nonumber \\
 &  & \left.\qquad\qquad+\left(\frac{1}{n}\sum_{\mathbf{x}_{\perp}}\text{Re }\text{tr}\left[U_{\boxdot}(\mathbf{x}_{\perp})\right]\right)^{N}+\left(\frac{1}{n}\sum_{\mathbf{x}_{\perp}}\text{Re }\text{tr}\left[U_{\boxdot}^{\dagger}(\mathbf{x}_{\perp})\right]\right)^{N}+\text{cross-terms}\right\} \nonumber \\
 & = & \frac{2}{N!}\Big(\frac{\beta}{2N}\Big)^{N}\left\{ n\mathcal{N}+\frac{\mathcal{N}_{\perp}}{n^{N}}+\text{cross-terms}\right\} .\label{eq:combi2}
\end{eqnarray}
This explains how the $\beta^{N}$ terms in (\ref{eq:zn2}) appear.
They become subleading for $N\geq3$.

With (\ref{eq:zn2}), we can take the combination 
\begin{equation}
\ln Z_{n}-n\ln Z_{1}=-\beta\mathcal{N}_{\perp}\left[\frac{1}{n}-n\right]+\frac{\beta^{2}\mathcal{N}_{\perp}}{4N^{2}}\left(\frac{1}{n^{2}}-n\right)+\frac{2}{N!}\Big(\frac{\beta}{2N}\Big)^{N}\left[n\mathcal{N}+\frac{\mathcal{N}_{\perp}}{n^{N}}\right]+O(\beta^{4}),\label{eq:combi}
\end{equation}
and then obtain the Renyi entropy
\begin{eqnarray}
S_{n} & = & \frac{\ln Z_{n}-n\ln Z_{1}}{1-n}\label{eq:snall}\\
 & = & -\beta\mathcal{N}_{\perp}\left(\frac{1+n}{n}\right)+\frac{\beta^{2}\mathcal{N}_{\perp}}{4N^{2}}\left(\frac{1+n+n^{2}}{n^{2}}\right)+\frac{2}{N!}\Big(\frac{\beta}{2N}\Big)^{N}\mathcal{N}_{\perp}\left(\frac{1}{n^{N}}-n\right)\frac{1}{1-n}+O(\beta^{4}).\nonumber 
\end{eqnarray}

This result implies, no matter how we place the lattice, as long as
the conical singularity is located inside the central plaquettes (i.e.
the central plaquettes exist), the result will be the same. But if
one chooses to put the conical singularity on a lattice site, then
there is no central plaquettes to offer the $\frac{1}{n}$ and $\frac{1}{n^{2}}$
factor in (\ref{eq:combi}), but the factors $n$ appears instead,
and the outcome of $\ln Z_{n}$ is canceled by that from $n\ln Z_{1}$,
and the Renyi entropy will vanish.

We express the Renyi entropy for the cases $N=2$, $N=3$, and $N>3$
explicitly,
\begin{eqnarray}
S_{n}^{(N=2)} & = & -\beta\mathcal{N}_{\perp}\frac{1+n}{n}+\frac{\beta^{2}\mathcal{N}_{\perp}}{2N^{2}}\left(\frac{1+n+n^{2}}{n^{2}}\right)+O(\beta^{4}),\label{eq:SN2}\\
S_{n}^{(N=3)} & = & -\beta\mathcal{N}_{\perp}\frac{1+n}{n}+\frac{\beta^{2}\mathcal{N}_{\perp}}{4N^{2}}\left(\frac{1+n+n^{2}}{n^{2}}\right)+\frac{\beta^{3}\mathcal{N}_{\perp}}{24N^{3}}\frac{(1+n)(1+n^{2})}{n^{3}}+O(\beta^{4}),\label{eq:SN3}\\
S_{n}^{(N>3)} & = & -\beta\mathcal{N}_{\perp}\frac{1+n}{n}+\frac{\beta^{2}\mathcal{N}_{\perp}}{4N^{2}}\left(\frac{1+n+n^{2}}{n^{2}}\right)+O(\beta^{4}).
\end{eqnarray}

As a result, the entanglement entropy of SU($N$) gauge theory in
the strong coupling expansion is 
\begin{eqnarray}
S_{\text{EE}}^{(N=2)} & = & \frac{A_{\perp}}{a^{2}}\left[-\frac{4N^{2}}{\lambda}+6\frac{N^{2}}{\lambda^{2}}+O(\frac{N^{4}}{\lambda^{4}})\right]+\delta S_{\text{EE}}^{(N=2)},\label{eq:result2}\\
S_{\text{EE}}^{(N=3)} & = & \frac{A_{\perp}}{a^{2}}\left[-\frac{4N^{2}}{\lambda}+3\frac{N^{2}}{\lambda^{2}}+\frac{4}{3}\frac{N^{3}}{\lambda^{3}}+O(\frac{N^{4}}{\lambda^{4}})\right]+\delta S_{\text{EE}}^{(N=3)},\label{eq:result3}\\
S_{\text{EE}}^{(N>3)} & = & \frac{A_{\perp}}{a^{2}}\left[-\frac{4N^{2}}{\lambda}+3\frac{N^{2}}{\lambda^{2}}+O(\frac{N^{4}}{\lambda^{4}})\right]+\delta S_{\text{EE}}^{(N>3)},\label{eq:result4}
\end{eqnarray}
where $A_{\perp}$ is the area of A, B region interface, $a$ is lattice
spacing, and $\lambda=g^{2}N$ is the t'Hooft coupling. As we argued
below Eq. (\ref{b}), this result is independent of how the lattice
is discretized, as long as the conical singularity is encircled by
the same number of plaquettes in the action. The additional terms
$\delta S_{\text{EE}}$ are the positive contribution from the cosmological
constant living on the 2-dimensional space transverse to the conical
singularity in the Lagrangian, which we expect to cancel the negative
term in the entanglement entropy, and will be explained in Sec. \ref{sec:2dCC}.

\section{\label{sec:2dCC}A 2-d Cosmological Constant Counterterm}

In the result Eqs. (\ref{eq:result2})$\sim$(\ref{eq:result4}),
one finds that the leading order $O(\beta)$ term in the entanglement
entropy has a negative contribution while all the subleading orders
are positive. It turns out that, with the 2 dimensional conical structure
in a 4 dimensional space time, we are allowed to introduce more local
operators in the continuum action 
\begin{equation}
S=\int d^{2}x_{\perp}d^{2}x_{\parallel}\left[-\frac{1}{4}F^{2}+c_{4}+c_{2}\delta^{(2)}(x_{\parallel},\tau)\right].\label{dS}
\end{equation}
The $c_{4}$ term is a 4-dimensional cosmological constant counter
term which does not contribute to the entanglement entropy. However,
the 2-dimensional cosmological constant counter term $c_{2}$, living
on the space transverse to the conical singularity and breaks the
translational symmetry on the cone, can contribute to the entanglement
entropy \cite{Cooperman:2013iqr}. Assuming $c_{2}$ is a smooth function
of $n$, then 
\begin{equation}
c_{2}=c_{2}^{^{\prime}}(n-1)+O\left((n-1)^{2}\right),
\end{equation}
where we $c_{2}^{^{\prime}}=c_{2}^{^{\prime}}(\beta)$ is a function
of $\beta$. We have made use of the fact that $c_{2}$ should vanish
at $n=1$ where translational symmetry is recovered. Therefore there
is an extra unspecified contribution to the entanglement entropy which
also obey the area law: 
\begin{equation}
\delta S_{\text{EE}}^{(N)}=A_{\perp}c_{2}^{^{\prime}(N)}.\label{eq:c2EE}
\end{equation}
We label the $N$ dependence explicitly since different theories would
have different $c_{2}^{^{\prime}}$ counter terms. Also, $c_{2}^{^{\prime}(N)}$
is $\beta$ dependent.

The negative leading term in our entanglement entropy could in principle
be compensated by the contribution from the 2-dimensional cosmological
constant in (\ref{eq:c2EE}). We remind the readers here that the
negative leading term arises from the constant term in the lattice
gauge field Lagrangian. Since the constant in the Lagragian for the
central plaquettes is different from that for the non-central ones
by a factor of $1/n$, the effect is just like the $c_{2}$ in Eq.(\ref{dS}).
If these constant terms are not included in the lattice Lagrangian,
like what was done in some of the actions studied previously \cite{Velytsky:2008rs,Gromov:2014kia,Aoki:2015bsa},
then the negative contribution to the entanglement entropy will not
arise, and the Lagrangian can not be reduced to the usual Yang-Mills
one at continuum limit. We argue that the $c_{2}$ and $c_{4}$ terms
will always appear by renormalization even they are set to zero at
certain renormalization scale. Setting them to be zero is equivalent
to choosing specific values for these counter terms. Previously it
was known that different choices of the boundary conditions gave different
values for the entanglement entropy \cite{Casini:2013rba}. This corresponds
to employing different regularization schemes, but those differences
can be compensated by having different values for the counter terms
for different regularization schemes used.

We further speculate that the 2-dimensional cosmological constant
can show up in the entanglement entropy of scalar and spinor field
theories as well. Also, they could play the role of a counter term
to absorb the ultraviolet divergence of entanglement entropy and make
entanglement entropy a finite physical quantity.

In spite of the discussion above, we would like to remind the reader
that, since there exist many possible center choices which may not
necessarily lead to the constant counter terms at the tip of the cone,
the inclusion of our proposed cosmological constants may not be sufficient
to achieve full agreement with the results from various center choices.

\section{Comparison with previous results}

Calculating the entanglement entropy of the guage fields is a subtle
problem because of the gauge invariance constraint and the nonlocality
of gauge fields on the lattice. Various approaches (\cite{Ghosh:2015iwa},
\cite{Donnelly:2014fua}-\cite{Radicevic:2015sza}) had been proposed
to tackle this problem, and \cite{Radicevic:2015sza} had pointed
out that these approaches are equivalent from the gauge invariant
local operator algebraic perspective \cite{Casini:2013rba}. \cite{Casini:2013rba}
elaborate the choice of the electric, magnetic, and trivial centers
in the operator algebra, corresponding to different choice of the
boundary condition for the local Hilbert space. The setup of our system,
as illustrated in Fig. \ref{fig:setup}, has the boundary cutting
across the links connecting the regions A and B. This configuration
is very similar to that in \cite{Buividovich:2008gq}, except that
ours has one more spatial dimension along the boundary plane. The
construction in \cite{Buividovich:2008gq} had been demonstrated in
\cite{Casini:2013rba} being equivalent to the electric center choice
in the algebraic approach, as they give rise to the same expectation
value to the same local operators. We remind the readers again that
the non-vanishing entanglement entropy in our model arises from the
existence of the central plaquette which encircles the conical singularity
on the Riemannian $n$-sheets in our the configuration, and this is
because we choose to place the boundary in between two adjacent lattice
sites, allow it to cut through the links connecting two different
regions.

Our result shows that the entanglement entropy between regions A and
B of $SU(N)$ Yang-Mills gauge fields on the lattice obeys the area
law up to the highest order $O(\beta^{3})$. At zero temperature and
large coupling, such system is expected to be confined. In the literature,
however, the area law is well known for the weak coupling theories
and conformal theories, and it is not yet well-established whether
the area law extends to the strong coupling and confining regime.
Our result in analytic formalism under the strong coupling expansion
suggests that the area law applies to the confining strong coupling
theories. This echoes \cite{Radicevic:2015sza} which demonstrates
the entanglement entropy of the strongly coupled $SU(N)$ lattice
gauge theory behaves as $|\partial V_{\perp}|\frac{\text{log }g^{2}N}{g^{8}N^{2}}$
at large $N$. On the other hand, \cite{Nakagawa:2009jk,Itou:2015cyu,Rabenstein:2018bri}
compute the $C$-function $C(l)=\frac{l^{3}}{|\partial A|}\frac{\partial S_{EE}}{\partial l}$
of the entanglement entropy for the Yang-Mills theory by lattice simulation
in $3+1$ dimensions, where $l$ is the size of the domain $A$ they
calculate the $S_{EE}$ for. The results show that for sufficiently
large $l$, $C(l)$ vanishes. Since $C(l)$ is normalized by the boundary
area $|\partial A|$ of the domain $A$, the results imply that the
entanglement entropy scales with the boundary area. The evidences
from lattice simulation also suggestion that the area law applies
to the confining Yang-Mills theories on the lattice, in accordance
with our result.

Our system is of infinite size in space and time dimensions. In \cite{Velytsky:2008rs},
the entanglement entropy of SU($N$) gauge fields on the lattice of
finite size at finite temperature is calculated in the Lagrangian
formalism by means of group characteristic expansions. In 1+1 dimensions,
when the periodic boundary condition is imposed on the spatial dimension,
the entanglement entropy is given by 
\begin{equation}
S_{\text{ent.}}=\left(\frac{\beta}{2N^{2}}\right)^{A/a^{2}}\left[1-\log\left(\left(\frac{\beta}{2N^{2}}\right)^{A/a^{2}}/N^{2}\right)\right]\label{S1}
\end{equation}
where $A$ is the total area of the 1+1 dimensional spacetime and
$a$ is the lattice spacing. At the limit of infinite area, the entanglement
entropy reduces to 0, instead of Eqs. (\ref{eq:result2})$\sim$(\ref{eq:result4}).
While the free boundary condition is chosen in the spatial dimensions,
the entanglement entropy vanishes identically. We can obtain the same
result if we set the conical singularity on a site so there is no
central plaquette at all. So the leading order result for a system
of periodic boundary condition in 1+1 dimensions comes from the configuration
with the plaquettes tiling the whole dimensions. But this configuration
gives vanishing entanglement entropy when the free spatial boundary
condition is taken. Despite this, the result in Eq.(\ref{S1}) does
not have the area law due to the nature of 1 spatial dimensional system,
and hence does not have the expected form.

Eqs. (\ref{eq:result2})$\sim$(\ref{eq:result4}) show the leading
UV-divergence $\sim\frac{A_{\perp}}{a^{2}}$ of the Yang-Mills entanglement
entropy in our setup, where the lattice spacing $a$ is the cutoff
of the system. It agrees with the leading divergence of $S_{EE}$
for the conformal theories in 3+1 dimensions. It would be interesting
to investigate whether the entanglement entropy of the Yang-Mills
fields (or even of other gauge theories, e.g. $Z_{2}$ gauge fields)
also exhibits logarithmic scaling in the subleading universal terms
in our setup, as our future work.

According to our result in Eq. (\ref{eq:result4}), the strongly coupled
$SU(N)$ Yang-Mills gauge theory at the large-$N$ limit in our setup
behave as $S_{\text{EE}}^{(N>>1)}=-\frac{A_{\perp}}{a^{2}}\frac{4N^{2}}{\lambda}=-\frac{A_{\perp}}{a^{2}}\left[\frac{4N}{g^{2}}\right]$.
In Section \ref{sec:2dCC}, we explained that this term could be compensated
by the contribution from the 2-dimensional cosmological constant in
(\ref{eq:c2EE}). The coefficient of this leading $S_{\text{EE}}^{(N>>1)}$
per unit entangling surface area vanishes if we allow $N$ and $g^{2}$
to approach infinity in such a way that $\frac{4N}{g^{2}}\to0$. We
can compare our result to that in \cite{Radicevic:2015sza}, where
the $S_{EE}$ of the strongly coupled $SU(N)$ lattice gauge theory
scales as $|\partial V_{\perp}|\frac{\text{log }(\lambda N)}{\lambda^{2}}\sim|\partial V_{\perp}|\frac{\text{log }(g^{2}N)}{g^{8}N^{2}}$
at large $N$, derived from the Hamiltonian approach. Note that \cite{Radicevic:2015sza}
sets lattice spacing to 1 and takes $\lambda\sim g^{4}N$, while in
our work we use $\lambda=g^{2}N.$ Since the replicating links and
the central plaquette prescription are not derivable from the Hamiltonian
formalism, it is not surprising that our aforementioned leading scaling
behavior in $S_{EE}$ does not agree with that in \cite{Radicevic:2015sza}.

Ref. \cite{Donnelly:2011hn} considers the entanglement entropy of
the ground states of the SU(2) Kogut-Susskind Hamiltonian \cite{Kogut:1974ag}
for the Wilson gauge theories at strong coupling limit, by including
the edge states living on the boundary into the Hilbert space, such
that the total entanglement entropy contains the contribution from
the edge states. For $d+1\geq3$ dimensions, the leading order entanglement
entropy obtained by \cite{Donnelly:2011hn} is 
\begin{equation}
S_{\text{ent.}}=\frac{A_{\perp}}{a^{2}}(d-1)\beta^{2}(\ln\frac{1}{\beta^{2}}+1+2\ln2)\label{S2}
\end{equation}
where $A_{\perp}$ is the boundary area \footnote{Note that the $\beta^{2}$ in \cite{Donnelly:2011hn} is equivalent
to our $\beta$. Eq. (\ref{S2}) is expressed in our notation.}. The entire leading contribution to the entropy in (\ref{S2}) is
given by the entanglement entropy of the edge modes, which corresponds
to the contribution of the cosmological constant counterterm in our
model. The non-local correlations of the d.o.f.'s in the two subregions
are manifest only at higher order. Ref. \cite{Buividovich:2008gq}
also gives similar result: it is demonstrated via numerical simulation
that, in the case of the $Z_{2}$ lattice gauge theory in 3 dimensions,
the entanglement entropy is almost saturated by the entropy of the
end points of the electric strings cut open by the boundary of the
two subregions. In our language, their setup corresponds to locating
the conical singularity on a site, so there is no central plaquette
at all such that the leading contribution is coming from the counterterm.

In our case, introducing the central plaquette incorporates the effect
of the edge modes on the boundary. In principle it is possible to
extract the dynamics of electric flux and magnetic flux on the boundary,
and the remaining contribution of the edge modes is naturally interpreted
as the cosmological constant counter-terms in the theory. This is
left for our future work.

\section{Summary}

To summarize, we have calculated the entanglement entropy of the $SU(N)$
Yang-Mills gauge theories on the lattice under the strong coupling
expansion in powers of $\beta=2N/g^{2}$. Using the replica method,
our Lagrangian formalism maintains gauge invariance on the lattice.
At $O(\beta^{2})$ and $O(\beta^{3})$, the entanglement entropy is
solely contributed by the central plaquettes enclosing the conical
singularity of the $n$-sheeted Riemann surface. The area law emerges
naturally to the highest order $O(\beta^{3})$ of our calculation.
The leading $O(\beta)$ term is negative, which could in principle
be canceled by taking into account the cosmological constant living
in interface of the two entangled subregions. This unknown cosmological
constant resembles the ambiguity of edge modes in the Hamiltonian
formalism. We have further speculated that this unknown cosmological
constant can show up in the entanglement entropy of scalar and spinor
field theories as well. Furthermore, it could play the role of a counterterm
to absorb the ultraviolet divergence of entanglement entropy and make
entanglement entropy a finite physical quantity. 

\begin{acknowledgments}
We would like to thank Michael Endres, David Lin, Feng-Li Lin, Masahiro
Nozaki, Chen-Te Ma, Jackson Wu and Yun-Long Zhang for helpful discussions.
This work is supported by the MOST, NTU-CTS and the NTU-CASTS of Taiwan.
JYC is supported by MOST Grant No. 105-2112-M-002-017-MY3 and No.
108-2112-M-002-003-MY3. SHD is supported by MOST Grant No. NSC103-2811-M-002-134.
JYP is supported in part by NSFC under grant No. 11125524 and 1221504.
\end{acknowledgments}

\bibliographystyle{unsrt}
\addcontentsline{toc}{section}{\refname}\nocite{*}
\bibliography{ref_ee}

\end{document}